\documentclass[letterpaper,10pt,conference]{ieeeconf}
\IEEEoverridecommandlockouts
\usepackage[T1]{fontenc}
\usepackage{amsmath,amssymb,graphicx,booktabs,multirow,tikz,cite,url,balance}
\usepackage{algorithm}
\usepackage{algpseudocode}
\usepackage{subcaption}
\usetikzlibrary{arrows,positioning}

\newtheorem{researchquestion}{Research Question}
\title{\LARGE\bf MATES: Learning Multi-Agent Interactions by Transforming Observations for Frozen Single-Agent Policies}
\author{Elie Abboud$^{1}$ and Oren Gal$^{2}$%
\thanks{$^{1}$Elie Abboud is with the department of Marine Technologies, University of Haifa, Israel.
Email: eliabboud1000@gmail.com.}%
\thanks{$^{2}$Oren Gal is with the Department of Marine Technologies, University of Haifa, Israel.
Email: orengal@univ.haifa.ac.il.}%
}

\begin{document}
\maketitle
\thispagestyle{empty}\pagestyle{empty}

\begin{abstract}

Multi-agent reinforcement learning (MARL) commonly trains decentralized policies from scratch, requiring agents to acquire individual task competence and coordination simultaneously. Yet many multi-agent problems admit a compatible single-agent counterpart in which the underlying task can be learned in isolation. We introduce Multi-Agent Observation Transformation for Existing Single-Agent Policies (MATES), an input-side adaptation framework for tasks whose multi-agent observations preserve the solo-task information while exposing separately identifiable neighbor information. From multi-agent experience, MATES learns a small adapter that maps this observation into the format expected by a frozen single-agent policy, inducing actions suited to the shared environment without updating the single-agent policy itself. MATES leaves the pretrained policy's internal architecture unchanged and retains the objectives and update procedures of the underlying MARL algorithm. We evaluate MATES using both on- and off-policy algorithms on lifelong pathfinding, navigation, and cooperative discovery, spanning discrete and continuous observation and action spaces. Across all evaluated settings, MATES optimizes only 3.5--7.3\% as many parameters as full-policy training while consistently outperforming MARL training from scratch. It approaches the performance of full fine-tuning, remains competitive overall with demonstration-based baselines, and retains strong task performance at team sizes not encountered during training. These results provide evidence that, under this observation structure, effective multi-agent behavior can be learned without modifying the policy that encodes individual competence.

\end{abstract}

\section{Introduction}
Deploying several capable agents together introduces a challenge that is not present when each acts alone: the actions of one agent change the conditions under which the others must succeed. During learning, this interdependence gives rise to non-stationarity, a central challenge in MARL \cite{marl-book}: as agents update their policies simultaneously in a shared environment, each must continually adapt to the changing behavior of the others. For example, in robot navigation (e.g. \cite{pogema, vmas}), a single agent may need only to navigate along a route toward its destination, whereas successful multi-agent navigation may additionally require avoiding collisions and coordinating with other agents \cite{CooperationMARL, pogema}.

A common approach to these problems is to train decentralized policies from scratch directly in the multi-agent environment, requiring them to acquire task-specific competence and coordination simultaneously \cite{marl-book}. For instance, Independent Proximal Policy Optimization \cite{mappo} and Independent Soft Actor-Critic (ISAC; \cite{sac, benchmarl}) train agents using experience generated while all agents interact simultaneously with the shared environment.

Many problems in the multi-agent setting have natural single-agent counterparts, in which the underlying individual task may be easier to learn. Nevertheless, standard MARL approaches that train decentralized policies from scratch do not attempt to utilize competence already acquired by an individual agent. For example, an agent trained to navigate a map by itself may already have learned to move toward its goal, follow a route, or search for a target. Training from scratch discards this existing capability and relearns both individual competence and coordination in the multi-agent environment. Although the effective dynamics and available observations change when other agents are introduced, we hypothesize that competence learned through single-agent interaction remains valuable. What changes is the context in which the resulting behavior must be expressed. This motivates our central research question:

\begin{researchquestion}
\label{rq:single-agent-transfer}
Can effective multi-agent behavior be learned without modifying a policy that already encodes individual competence?
\end{researchquestion}

We investigate this question in tasks whose observation structure reflects the distinction between individual competence and multi-agent coordination. Let $o_i$ denote the observation available to agent $i$ in the single-agent setting. When the agent is deployed with teammates, its observation expands to include local information $N_i$ about neighboring agents, yielding $(o_i,N_i)$. Thus, $o_i$ contains the information on which the agent originally learned its individual task competence, while $N_i$ provides the additional context needed to respond appropriately to teammates. This decomposition provides a natural interface for adapting the agent's behavior to the multi-agent setting while preserving the policy that encodes its individual competence.

To exploit this interface, we introduce Multi-Agent Observation Transformation for Existing Single-Agent Policies (MATES), an input-side adaptation framework. A small adapter receives $(o_i,N_i)$ and produces a transformed observation $\tilde{o}_i$ in the format expected by the frozen single-agent policy, which we denote by $\pi_{\mathrm{solo}}$. The policy $\pi_{\mathrm{solo}}$ is trained exclusively in the single-agent setting and remains frozen upon transfer to the multi-agent setting. MATES uses multi-agent experience to optimize its lightweight trainable components while leaving the action-producing single-agent policy unchanged. It retains the objectives and update procedures of the underlying MARL algorithm, augmenting only the architecture with lightweight components and restricting optimization to the designated trainable components. MATES thereby learns how the newly available neighbor information should transform the observation presented to $\pi_{\mathrm{solo}}$. Because this adaptation does not require modifying the internal architecture of $\pi_{\mathrm{solo}}$, MATES is compatible with a broad range of policy architectures. At execution time, each agent independently applies the shared adapter--policy composition to its own local observation, without relying on a centralized controller.

Our contributions are as follows:

\begin{enumerate}
\renewcommand{\labelenumi}{(\roman{enumi})}

\item We introduce MATES, a general framework for transferring a competent single-agent policy to multi-agent settings through lightweight observation adaptation. For tasks with a compatible single-agent counterpart, MATES can be integrated with MARL algorithms without changing their learning objectives or the internal architecture of the pretrained policy.

\item We evaluate MATES on three tasks---lifelong pathfinding, navigation, and cooperative discovery---spanning discrete and continuous observation and action spaces. We instantiate its lightweight adapter as either a CNN or an MLP according to the observation modality and integrate it with both on-policy and off-policy MARL algorithms. Across all evaluated settings, MATES optimizes only $3.5$--$7.3\%$ as many parameters as full-policy training while consistently outperforming MARL training from scratch.

\item We show that MATES approaches the performance of full fine-tuning while updating substantially fewer parameters and remains competitive overall with PegMARL and R2BC. Whereas PegMARL and R2BC transfer prior behavior through demonstrations, MATES transfers competence by directly retaining the pretrained single-agent policy as a frozen controller, without requiring a demonstration buffer or imitation-learning updates.

\item We demonstrate that MATES generalizes beyond its training team size, retaining strong task performance across agent counts not encountered during training.

\end{enumerate}

\section{Related Work}

\subsection{Transfer and Demonstration-Guided Multi-Agent Learning}

Transfer learning in reinforcement learning seeks to improve learning in a target task by reusing knowledge acquired from related source tasks. This knowledge can be transferred through policy reuse, policy distillation, value functions, experience, models, and other learned representations; broader treatments of transfer in RL and MARL are provided by \cite{taylor2009transfer,dasilva2019transfer}.

Imitation learning (IL) seeks to reproduce an expert policy from demonstrations generated by an expert. Behavior cloning (BC) is a commonly employed IL technique that estimates the expert policy through supervised learning on expert demonstrations \cite{bojarski2016end,BC4AC}. Round-Robin Behavior Cloning (R2BC) \cite{r2bc} extends BC to multi-agent systems by allowing an expert to control one agent at a time while the remaining agents execute their current policies. The resulting per-agent demonstrations are accumulated and used to update the policies through BC.

Beyond direct imitation, demonstrations can also guide reinforcement learning. In MARL, Personalized Expert-Guided MARL (PegMARL) \cite{pegmarl} uses personalized single-agent demonstrations to train behavior and transition discriminators whose outputs provide reward-shaping signals for learning a multi-agent policy from scratch.

R2BC and PegMARL therefore rely on demonstration data and imitation or demonstration derived learning signals. MATES uses neither: it preserves the pretrained single-agent policy itself as a frozen, action-producing controller and learns its observation-side adaptation directly from multi-agent experience.

\subsection{Adaptation of Frozen Models and Policies}

Parameter-efficient adaptation repurposes pretrained models by optimizing a restricted set of parameters rather than fine-tuning the complete model. Common approaches include adapter modules, low-rank updates, and prompt-based methods \cite{ding2023parameter,han2024parameter}. Adapter tuning inserts trainable modules within a frozen network, whereas side-tuning combines the output of a lightweight side network with that of the unchanged pretrained network \cite{houlsby2019parameter,zhang2020sidetuning}. MATES instead leaves the internal computation of the pretrained policy unchanged and adapts the observation presented to it.

Related policy-adaptation methods preserve an existing policy or controller while learning an additional component. Residual reinforcement learning adds a learned correction to the action of a fixed controller \cite{silver2018residual,johannink2019residual}; MATES instead adapts the observation presented to the controller without directly correcting its actions. VR-Goggles for Robots transforms real-world images into the simulated visual domain expected by a pretrained policy \cite{zhang2019vrgoggles}; MATES likewise operates at the policy's input, but seeks to adapt its behavior to multi-agent interaction rather than retain its learned behavior across a domain shift.

\section{Preliminaries}

\subsection{Decentralized Cooperative Multi-Agent Reinforcement Learning}

We model the task as a partially observable Markov game with $n$ homogeneous agents:

$$
\mathcal G=(\mathcal S,\mathcal A^n,\bar{\mathcal O}^n,P,\Omega,\{r_i\}_{i=1}^n,\rho_0,\gamma, n).
$$

At timestep $t$, state $s_t \in \mathcal S$ yields each agent a local observation $\bar o_{i,t}\sim\Omega_i(\cdot\mid s_t)$. Each agent $i$ samples $a_{i,t}\sim\pi_\theta(\cdot\mid\bar o_{i,t})$, forming the joint action $\mathbf a_t=(a_{1,t},\ldots,a_{n,t})$. The environment draws $s_{t+1}\sim P(\cdot\mid s_t,\mathbf a_t)$ and assigns $r_i(s_t,\mathbf a_t,s_{t+1})$. We consider the setting in which agents share parameters. The objective is

$$
J(\pi_\theta)=\mathbb E_{\tau\sim\pi_\theta}\!\left[\sum_{t=0}^{T-1}\gamma^t\sum_{i=1}^n r_i(s_t,\mathbf a_t,s_{t+1})\right].
$$

Under decentralized training and decentralized execution (DTDE) we consider, neither the actors nor the critics receive the global state or joint action.

\begin{figure*}[!t]
    \centering
    \includegraphics[width=\textwidth]{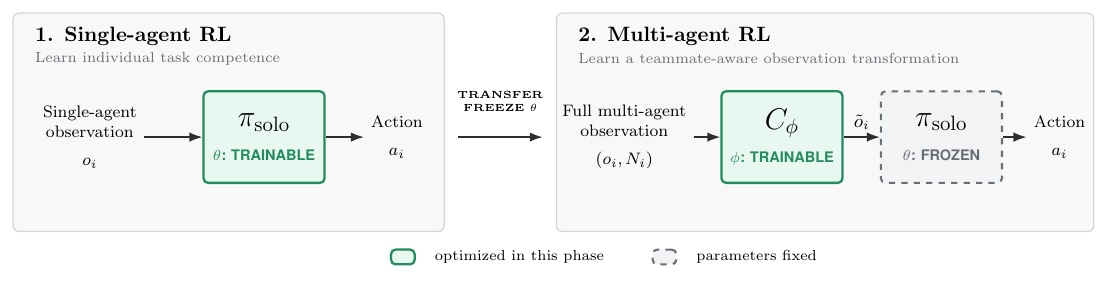}
    \caption{\textbf{Overview of MATES.} Training proceeds in two sequential phases. First, the parameters $\theta$ of a single-agent policy $\pi_{\mathrm{solo}}$ are trained normally using the single-agent observation $o_i$. Second, $\pi_{\mathrm{solo}}$ is transferred to the multi-agent setting and $\theta$ is frozen, while the parameters $\phi$ of an observation adapter $C_\phi$ are trained using the underlying MARL method to map the full multi-agent observation $\bar{o}_i=(o_i,N_i)$ to a transformed observation $\tilde{o}_i$ consumed by $\pi_{\mathrm{solo}}$. Solid green and dashed gray boxes denote trainable and frozen components, respectively. The figure depicts the action-selection pathway; critic-side components are omitted for clarity.}
    \label{fig:mates-overview}
\end{figure*}

\subsection{Single-Agent Counterpart and Frozen Policy}

We consider tasks with a compatible single-agent counterpart sharing action space $\mathcal A$. Its observation $o_i\in\mathcal O_{\mathrm{solo}}$ retains the same task semantics after deployment, while the multi-agent observation is $\bar o_i=(o_i,N_i)$. Here, $N_i$ locally summarizes neighboring agents and may have fixed size rather than enumerate the team. A solo policy $\pi_{\mathrm{solo}}(a_i\mid o_i)$ is trained in this counterpart. During transfer, it is frozen: its parameters are not updated by multi-agent training, although gradients may pass through it to optimize components preceding its input.

We focus on problems admitting the explicit decomposition $\bar{o}_i=(o_i,N_i)$ to maintain a clear separation between task-specific information and multi-agent interactions. This allows our experiments to directly evaluate whether MATES maps the multi-agent extension of the problem into the competency space of the frozen solo policy. The architecture itself does not require this decomposition and could, in principle, be applied whenever a compatible single-agent counterpart exists, although we do not evaluate that broader setting.

\section{Method}
\label{sec:method}

We now present MATES. As illustrated in Figure~\ref{fig:mates-overview}, the method proceeds in two sequential phases. First, a policy $\pi_{\mathrm{solo},\theta}$ is trained in the single-agent counterpart to acquire task-specific competence. MATES then transfers this policy unchanged to the multi-agent task and freezes its parameters $\theta$. Because multi-agent observation additionally provides information about neighboring agents that $\pi_{\mathrm{solo},\theta}$ was not trained to process, MATES places a trainable observation adapter $C_\phi$ before the frozen policy. For each agent $i$, the adapter maps the full multi-agent observation $\bar o_i=(o_i,N_i)$ to a solo-compatible observation

\begin{equation}
\tilde o_i=C_\phi(o_i,N_i),
\qquad \tilde o_i\in\mathcal O_{\mathrm{solo}}.
\end{equation}

The resulting MATES policy is the composition

\begin{equation}
\pi_{\mathrm{MATES},\phi}(a_i\mid\bar o_i)
=
\pi_{\mathrm{solo},\theta}
\bigl(a_i\mid C_\phi(\bar o_i)\bigr).
\end{equation}

Thus, neighboring-agent information affects action selection by changing the effective observation of the solo policy rather than by modifying or replacing its learned controller.

During the multi-agent phase, $\phi$ is optimized using the objective of the underlying MARL algorithm.

Although $\pi_{\mathrm{solo},\theta}$ is frozen, it remains part of the differentiable computation graph. Gradients therefore pass through the policy with respect to its input and update $C_\phi$, while no gradient updates are applied to $\theta$. Consequently, multi-agent experience teaches the adapter which transformations of $(o_i,N_i)$ elicit effective behavior from the preserved single-agent controller. See Algorithm~\ref{alg:mates}, describing MATES in psuedocode.

\begin{algorithm}[t]
\caption{Multi-Agent Observation Transformation for Existing Single-Agent Policies (MATES)}
\label{alg:mates}
\begin{algorithmic}[1]

\Require Compatible single-agent task $\mathcal G_{\mathrm{solo}}$ and multi-agent task $\mathcal G_{\mathrm{MA}}$

\State $\pi_{\mathrm{solo},\theta} \gets
\Call{RL-Train}{\mathcal G_{\mathrm{solo}}}$

\State Freeze $\theta$ and initialize the observation adapter $C_\phi$

\State Define $\pi_{\mathrm{MATES},\phi}(a_i\mid\bar o_i)
= \pi_{\mathrm{solo},\theta}
\bigl(a_i\mid C_\phi(\bar o_i)\bigr)$

\State $\phi^\star \gets
\Call{MARL-Train}{
\mathcal G_{\mathrm{MA}},
\pi_{\mathrm{MATES},\phi};
\theta\ \mathrm{frozen}
}$

\State \Return $\pi_{\mathrm{MATES},\phi^\star}$

\end{algorithmic}
\end{algorithm}

\subsection{Implementation Variants}

MATES does not require the observation adapter to follow a particular parameter-sharing scheme. A shared adapter $C_\phi$ may be applied to every agent, as in Algorithm~\ref{alg:mates}, or each agent may have an independently parameterized adapter $C_{\phi_i}$. Sharing is natural for homogeneous agents and keeps the number of trainable parameters independent of the team size, whereas independent adapters permit agent-specific transformations. We employ shared adapters in our experiments.

Actor--critic methods additionally require the critic to operate under the multi-agent observation distribution. The critic learned during single-agent training is valuable for the same reason as the solo policy: it has already learned to evaluate progress in the underlying task. We therefore transfer and freeze the solo critic and place a separately initialized observation adapter before it. For a value-function critic, the actor and critic become

\begin{subequations}
\renewcommand{\theequation}{\theparentequation.\arabic{equation}}
\begin{align}
\pi_{\mathrm{MATES}}(a_i\mid\bar o_i)
&=
\pi_{\mathrm{solo},\theta_\pi}
\bigl(a_i\mid C^\pi_{\phi_\pi}(\bar o_i)\bigr),\\
V_{\mathrm{MATES}}(\bar o_i)
&=
V_{\mathrm{solo},\theta_V}
\bigl(C^V_{\phi_V}(\bar o_i)\bigr),
\end{align}
\end{subequations}

where $C^\pi_{\phi_\pi}$ and $C^V_{\phi_V}$ are distinct adapters and $\theta_\pi$ and $\theta_V$ remain frozen. For an action-value critic, the corresponding construction is

\begin{equation}
Q_{\mathrm{MATES}}(\bar o_i,a_i)
={}
Q_{\mathrm{solo},\theta_Q}
\bigl(C^Q_{\phi_Q}(\bar o_i,a_i\bigr)).
\end{equation}

The underlying actor and critic objectives optimize their respective adapters. When multiple critics are used, each critic receives its own independently initialized adapter.

We also consider optionally training the transferred critic's output head while keeping the remainder of the critic frozen. This allows the final value prediction to be recalibrated for the multi-agent task without relearning the critic's internal representation. We refer to the variant that trains this head as MATES and to the variant that keeps it frozen as MATES-NH\footnote{MATES-NH expands to MATES-No Head trained.}. As we will see in our experiments, this additional flexibility has no interpretible effect on performance.

\section{Experiments}
\label{sec:experiments}

\subsection{Research Questions}

We organize our evaluation around four research questions:

\begin{itemize}
\item[\textbf{RQ1}] \textbf{Performance and parameter efficiency:} Can MATES attain competitive multi-agent task performance while optimizing only a small fraction of the parameters of a conventional MARL policy during transfer?

\item[\textbf{RQ2}] \textbf{Generalization across team sizes:} Does an adapter trained at one team size remain effective when deployed with different, including substantially larger, numbers of agents?

\item[\textbf{RQ3}] \textbf{Source of performance:} Does MATES benefit specifically from transferring a competent solo policy, or can comparable performance be obtained from the compact adapter architecture alone?

\item[\textbf{RQ4}] \textbf{Value-head adaptation:} Is training the transferred critic's value head necessary, or is adapting its observation sufficient?

\end{itemize}

\subsection{Tasks and Metrics}

We evaluate MATES on three partially observable multi-agent tasks spanning one discrete grid navigation environment and two continuous multi-robot control.

\textbf{POGEMA. \cite{pogema}}  is a grid-based multi-agent pathfinding environment in which agents navigate toward assigned goals while avoiding obstacles and one another. Upon reaching a goal, an agent is assigned another, requiring policies to sustain navigation over time. We measure task completion using \emph{throughput}, defined as the number of goals reached per environment step; higher values indicate better performance.

\textbf{Navigation. \cite{vmas}} The VMAS Navigation task places continuous-control agents at randomly sampled positions and assigns each agent a goal. Agents must reach their respective goals while avoiding collisions, using local state and range-sensor observations. When all agents simultaneously reach their goals, the episode ends. We measure task success as the number of steps required to complete an episode; lower values indicate faster task completion.

\textbf{Discovery. \cite{vmas}} In VMAS Discovery, agents must locate and cover targets while avoiding collisions. Covered targets are respawned, requiring agents to continually redistribute themselves throughout the environment. We report the number of targets covered during an episode; higher values indicate better performance.

These metrics directly measure completion of the underlying task rather than the rewards used for optimization.

\subsection{Learning Algorithms}

We instantiate MATES with Independent Proximal Policy Optimization (IPPO) \cite{mappo} and Independent Soft Actor--Critic (ISAC) \cite{sac}, as implemented in BenchMARL \cite{benchmarl}. IPPO is an on-policy actor--critic method based on PPO, whereas ISAC is an off-policy, entropy-regularized actor--critic method based on SAC. Evaluating both allows us to test whether observation-side transfer remains effective under distinct optimization and data-reuse regimes.

We evaluate POGEMA with IPPO and evaluate Navigation and Discovery with both IPPO and ISAC. In every setting, the solo policy and corresponding conventional MARL policy use the same network architecture. The solo policy is trained using the single-agent version of the same learning algorithm, while multi-agent policies share parameters across homogeneous agents.

\begin{figure*}[t]
    \centering

    \begin{subfigure}{\textwidth}
        \centering
        \includegraphics[
            width=\textwidth,
            height=0.39\textheight,
            keepaspectratio
        ]{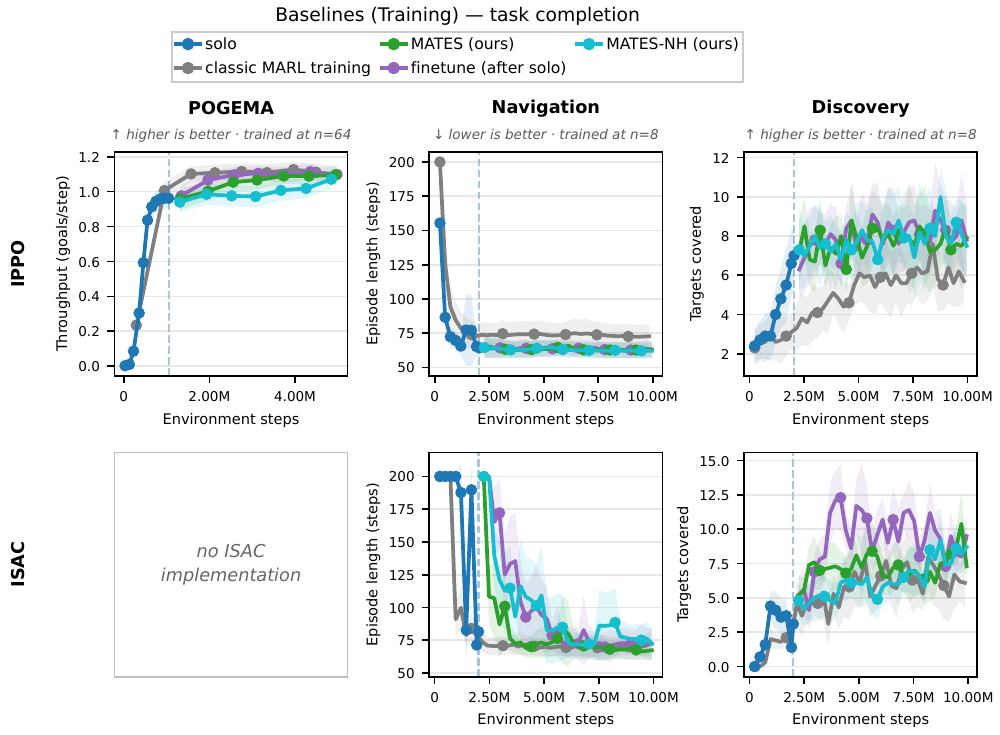}
        \caption{Direct reinforcement-learning baselines.}
        \label{fig:training-direct-baselines}
    \end{subfigure}

    \vspace{2mm}

    \begin{subfigure}{\textwidth}
        \centering
        \includegraphics[
            width=\textwidth,
            height=0.39\textheight,
            keepaspectratio
        ]{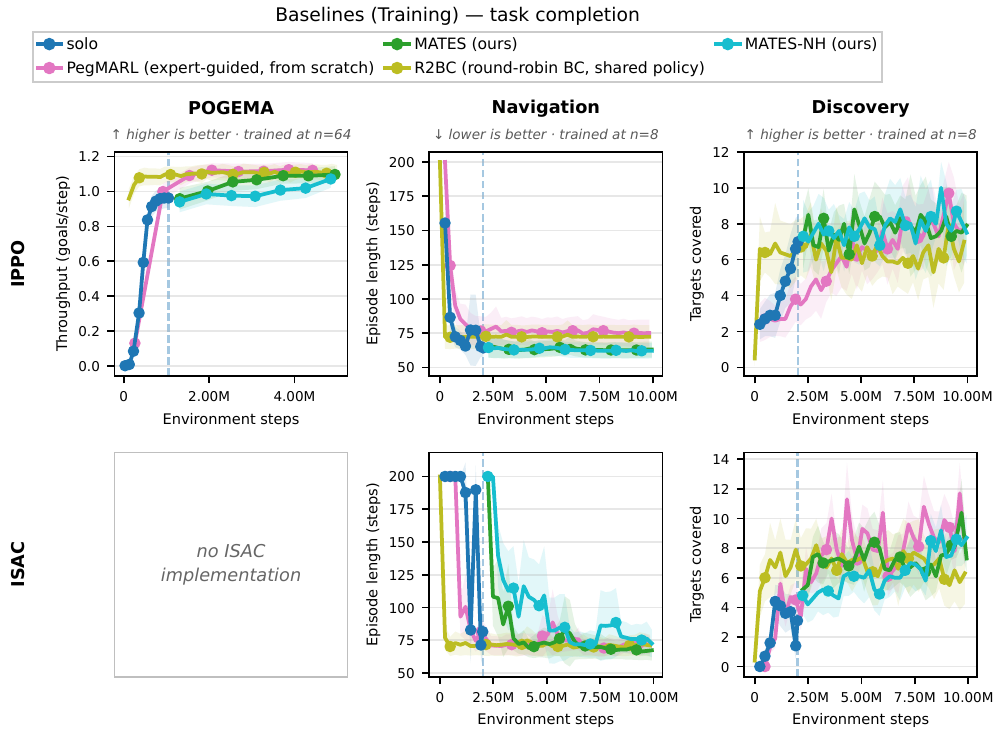}
        \caption{Expert-guided baselines.}
        \label{fig:training-expert-baselines}
    \end{subfigure}

    \caption{\textbf{Task completion during training.}
    MATES and MATES-NH are compared with
    (\subref{fig:training-direct-baselines}) direct reinforcement-learning
    baselines and (\subref{fig:training-expert-baselines}) expert-guided
    baselines. Dashed vertical lines mark the transition from solo
    pretraining to multi-agent training. Curves and shaded regions report
    the mean and 95\% confidence interval over 30 held-out evaluation
    episodes, respectively.}
    \label{fig:training-baselines}
\end{figure*}

\begin{figure*}[t]
    \centering
    \includegraphics[width=0.7\textwidth]
    {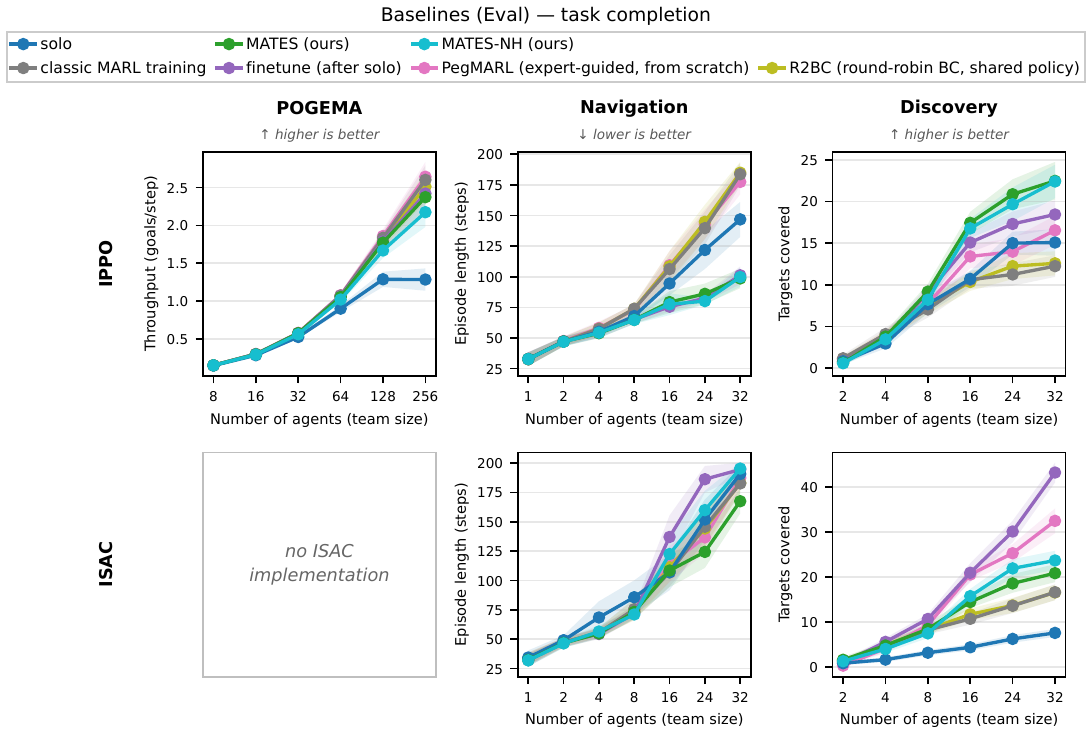}
    \caption{\textbf{Generalization across team sizes.}
    Policies trained with 64 agents in POGEMA and 8 agents in Navigation
    and Discovery are evaluated without further training at the team sizes
    shown. Points and shaded regions report the mean and 95\% confidence
    interval over 30 held-out episodes, respectively.}
    \label{fig:team-size-baselines}
\end{figure*}

\begin{table*}[t]
\centering
\caption{\textbf{Parameters optimized during multi-agent training.}
Percentages are relative to the corresponding fully trainable network.
MATES-NH denotes MATES without value-head training.}
\label{tab:trainable-parameters}
\footnotesize
\setlength{\tabcolsep}{7pt}
\begin{tabular}{@{}lccccc@{}}
\toprule
\textbf{Method} &
\textbf{POGEMA} &
\shortstack{\textbf{Navigation}\\\textbf{IPPO}} &
\shortstack{\textbf{Discovery}\\\textbf{IPPO}} &
\shortstack{\textbf{Navigation}\\\textbf{ISAC}} &
\shortstack{\textbf{Discovery}\\\textbf{ISAC}} \\
\midrule
Classic MARL
& 121,542 (100\%)
& 142,602 (100\%)
& 149,258 (100\%)
& 214,538 (100\%)
& 224,522 (100\%) \\
MATES
& 8,931 (7.3\%)
& 5,029 (3.5\%)
& 8,383 (5.6\%)
& 8,189 (3.8\%)
& 13,220 (5.9\%) \\
MATES-NH
& 1,186 (1.0\%)
& 4,772 (3.3\%)
& 8,126 (5.4\%)
& 7,675 (3.6\%)
& 12,706 (5.7\%) \\
\bottomrule
\end{tabular}
\end{table*}

\subsection{Methods and Baselines}

\textbf{MATES variants.}
We evaluate MATES together with MATES-NH, its no-head variant. Both methods freeze the transferred solo actor and critic and train observation adapters during the multi-agent phase. MATES additionally trains the critic's value head, whereas MATES-NH keeps this head frozen. Their precise architectures and numbers of trainable parameters are reported in Section \ref{sec:architectures-parameters}.

\textbf{Direct training and transfer baselines.}
We compare against three direct baselines. \emph{Solo} deploys the pretrained single-agent policy independently for every agent, without access to neighboring-agent information or subsequent multi-agent training. \emph{Classic MARL} trains the complete parameter-shared policy from random initialization directly on the multi-agent task. \emph{Full fine-tuning} expands the input layer of the pretrained solo network to accept the multi-agent observation, transfers the pretrained weights and optimizes the complete network during multi-agent training.

\textbf{Expert-guided baselines.} We additionally compare against PegMARL \cite{pegmarl} and R2BC \cite{r2bc}, both of which learn from expert demonstrations. PegMARL trains behavior and transition discriminators on demonstrations from the same solo expert used by MATES and uses their outputs to shape the agents' rewards. Its original formulation employs agent-specific discriminators and policies together with a shared team reward. We instead share the discriminators and policy parameters across homogeneous agents and apply the shaping signal to the environments' native per-agent rewards. Besides matching the setting used by all other methods, parameter sharing permits the resulting policy to be evaluated at team sizes different from the training team size.

R2BC is a behavior-cloning baseline in which a frozen expert controls agents in round-robin order to provide online action labels for a newly initialized learner. We use the corresponding Classic MARL policy as the expert, rather than the MAPPO expert used in the original implementation, and train a single parameter-shared actor to reflect the homogeneous-agent setting.

\begin{figure*}[t]
    \centering
    \includegraphics[width=0.7\textwidth]
    {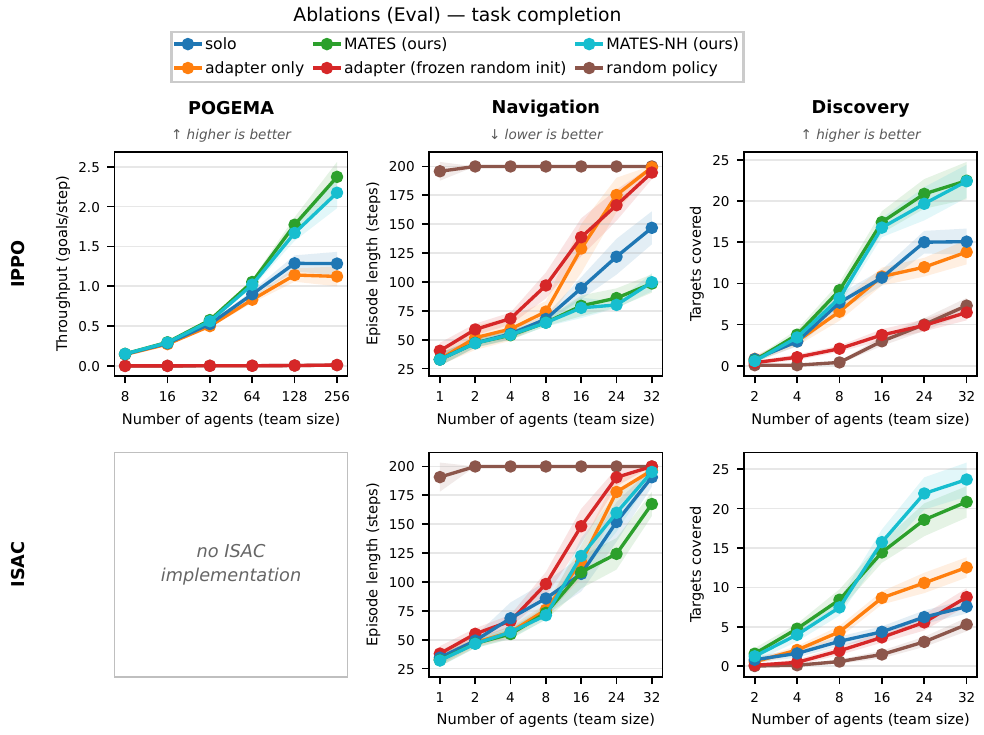}
    \caption{\textbf{Ablation results across team sizes.}
    Points and shaded regions report the mean and 95\% confidence interval
    over 30 held-out evaluation episodes, respectively.}
    \label{fig:team-size-ablations}
\end{figure*}

\subsection{Architectures and Parameter Efficiency} \label{sec:architectures-parameters}

All experiments use decentralized, parameter-shared networks: one network is
shared by every homogeneous agent, and observation dimensionality does not
grow with team size.

For Navigation and Discovery, we use the standard BenchMARL
\cite{benchmarl} backbones: IPPO uses separate two-layer width-256 actor and
critic MLPs, while ISAC uses the same actor together with twin two-layer
width-256 Q-networks. For POGEMA, we follow the backbone architecture of
Follower \cite{follower}, using a shared 64-channel ResNet encoder
with one residual block. CNN components use ReLU activations, whereas the
VMAS MLPs use Tanh.

The MATES adapters are substantially smaller than these backbones. POGEMA
uses a two-layer CNN adapter with 32 hidden channels. In VMAS, each adapter
is a one-hidden-layer width-64 MLP: IPPO uses independent adapters for its
actor and critic, while ISAC uses one actor adapter and an independent
joint observation--action adapter for each Q-network.

As specified in Algorithm~\ref{alg:mates}, MATES freezes the transferred
single-agent parameters \(\theta\) before multi-agent training and optimizes
only the newly introduced adapter parameters \(\phi\). We apply the same
principle to the transferred critic. In the IPPO instantiation, MATES also
updates the critic's existing value head, whereas MATES-NH keeps this head
frozen and trains only the adapters.

Table~\ref{tab:trainable-parameters} quantifies the parameter efficiency of
MATES by comparing the number of parameters optimized during multi-agent
training with full-network optimization. Following
Algorithm~\ref{alg:mates}, MATES freezes the transferred backbone and trains
only the adapters and, where applicable, the critic's value head.

Relative to the parameter count of the corresponding full MARL network, the
number of parameters optimized by MATES is \(3.5\%\)--\(7.3\%\), while that
of MATES-NH is \(1.0\%\)--\(5.7\%\) across all evaluated settings. Thus, MATES requires a substantially smaller trainable parameter
budget than full-network optimization.

\subsection{Results}

\textbf{Performance, parameter efficiency, and generalization.}

Figure~\ref{fig:training-baselines} compares task performance throughout
training. MATES and MATES-NH perform competitively with Classic MARL, full
fine-tuning, PegMARL, and R2BC across the evaluated tasks and learning
algorithms. No baseline consistently dominates across all settings.

This performance does not require optimizing the transferred backbone.
As shown in Table~\ref{tab:trainable-parameters}, the trainable parameter
count of MATES is equivalent to only \(3.5\%\)--\(7.3\%\) of the
corresponding full-network parameter count.

Figure~\ref{fig:team-size-baselines} evaluates the final policies across
team sizes without additional training. Although MATES is trained with 64
agents in POGEMA and 8 agents in Navigation and Discovery, its performance
remains competitive as the number of agents increases. Its advantage over
the unadapted solo policy generally widens beyond the training team size,
and MATES frequently matches or exceeds the fully trained baselines. This is
particularly evident in Navigation and Discovery under IPPO and in
Navigation under ISAC. Together, these results answer RQ1 affirmatively:
MATES achieves competitive multi-agent performance with a substantially
smaller trainable parameter budget. They also answer RQ2 affirmatively:
the learned adaptation remains effective beyond the team size encountered
during training.

\textbf{Contributions of solo competence and observation adaptation.}

Figure~\ref{fig:team-size-ablations} separates the contributions of the
transferred solo policy and the learned adapter. The solo policy successfully
completes each task and substantially outperforms the random policy,
confirming that it transfers useful task competence. However, its relative
performance deteriorates as team size increases, and the gap between the
solo policy and MATES generally widens.

The adapter-only policy also learns nontrivial behavior but consistently
underperforms MATES. When the pretrained solo network is instead replaced
by a frozen randomly initialized network, the trained adapter can recover
limited performance in some settings, but it does not approach MATES and
degrades toward random-policy performance as team size increases. Thus,
an arbitrary frozen backbone is insufficient. Together, these results answer
RQ3: MATES derives its performance from combining a competent pretrained
policy with an adapter that accommodates multi-agent interaction, rather
than from either component alone.

\textbf{Value-head adaptation.} Under IPPO, MATES and MATES-NH perform similarly in Navigation and
Discovery, indicating that observation adaptation is generally sufficient
without updating the transferred value head. Updating the value head provides
a slight benefit in POGEMA, particularly at larger team sizes, while having opposite roles across ISAC in the evaluated settings. RQ4
therefore has a qualified answer: value-head adaptation is not generally
necessary, although it can improve performance in some settings.

\section{Conclusion}

We introduced MATES, a parameter-efficient approach for transferring
single-agent policies to multi-agent tasks. MATES preserves the pretrained
actor and critic while learning compact input-side adapters that account for
multi-agent interaction. Across both discrete and continuous environments instantiated with on-policy IPPO
and off-policy ISAC, MATES performs competitively with full MARL training, full
fine-tuning, and expert-guided baselines while optimizing a parameter count
equivalent to only \(3.5\%\)--\(7.3\%\) of the corresponding full-network
size. The learned adapters also remain effective when deployed at team sizes
not encountered during training. Our ablations show that these results
depend on both components of the method: solo pretraining supplies useful
task competence, while observation adaptation enables that competence to
scale to unseen increasing counts of multi-agent deployment. Critic-head adaptation provides
task-dependent benefits but is not consistently required.

\section*{Acknowledgment}
OpenAI ChatGPT assisted with framing and language in this work, and Cursor (mainly Claude and Grok) assisted with generating the code, edit visualization of the plots, verifying implementation and end-to-end correctness via comprehensive testing. The ideas, algorithms, experiments, results, and implementation are entirely the authors. The authors retain responsibility for all content.

\balance
\bibliographystyle{ieeetr}
\bibliography{references}
\end{document}